\documentclass{ws-ijmpa}

\usepackage{amsmath}
\usepackage{amssymb}
\usepackage{overcite}

\newcommand{\half}{\frac{1}{2}}

\newcommand{\pdho}[2]{\ensuremath{\frac{\partial^{#1} }
        {\partial #2^{#1} }}}

\newcommand{\slashletter}[1]{\ensuremath{\kern+0.1em /\kern-0.65em #1}}

\begin{document}

\setcounter{secnumdepth}{1}

\markboth{M. P. FRY}{Fermion Determinants: Some Recent Analytic Results}

\title{FERMION DETERMINANTS: SOME RECENT ANALYTIC RESULTS}

\author{\footnotesize M. P. FRY}

\address{School of Mathematics, University of Dublin\\
Dublin 2, Ireland}

\maketitle

\begin{abstract}
The use of known analytic results for the continuum fermion
determinants in QCD and QED as benchmarks for zero lattice spacing
extrapolations of lattice fermion determinants is
proposed. Specifically, they can be used as a check on the
universality hypothesis relating the continuum limits of the na\"{\i}ve,
staggered and Wilson fermion determinants.
\end{abstract}

\section{Introduction}
\label{Sec_01}

Without internal quark loops such basic phenomena as color charge
screening and fast quark fragmentation into hadrons cannot be
formulated. Accordingly, the fermion determinant is one of the central
quantities in any \emph{ab initio} lattice QCD calculation that will confront
experiment.

It is proposed to use analytic results for fermion determinants
in QCD and QED as a check on extrapolated lattice fermion determinants.
In particular, the analytic results should be used to verify the
universality hypothesis relating the continuum limits of the na\"{\i}ve,
staggered and Wilson fermion determinants for a single quark
flavor~\cite{A_Adams04}:

\large

\begin{eqnarray*}
(\text{det}_{\text{Wilson}})^{16} &\underset{a \to 0}{=}& \text{det}_{\text{na\"{\i}ve}}\\
(\text{det}_{\text{Wilson}})^{4} &\underset{a \to 0}{=}& \text{det}_{\text{staggered}},
\end{eqnarray*}

\normalsize

\noindent
or stated somewhat differently, verify that

\large

\begin{displaymath}
\text{det}_{\text{Wilson}},
(\text{det}_{\text{staggered}})^{\frac{1}{4}},
(\text{det}_{\text{na\"{\i}ve}})^{\frac{1}{16}} \underset{a \to 0}{=}
  \text{det}_{\text{analytic}}.
\end{displaymath}

\normalsize

\noindent
It is understood that these are limits computed by taking ratios of
the interacting and free determinants. Checks such as these would also
be useful in establishing error bounds on determinant extrapolations
to the continuum. One important point should be noted:
$\text{det}_{\text{analytic}}$ is renormalization scheme dependant,
and so it is necessary to have compatible lattice and continuum
renormalization schemes when making the above comparisons.

The analytic results, to be relevant and useful, should be free of
ultraviolet and volume cutoffs. Since large quark mass can be dealt
with by perturbation theory, the results should be valid for all mass
values or at least mass values small enough to connect with effective
chiral theories. Finally, the background gauge fields should
be as general as possible. Obviously these are severe constraints that
greatly limit the available benchmark determinants. For lattice
theorists such checks will require writing new code and convincing
funding agencies that this is a worthwhile project.

\section{Defining the Fermion Determinant}
\label{Sec_02}

By fermion determinant we mean the ratio of determinants of the
interacting and free Euclidean Dirac operators,
$\text{det}(\slashletter{P} - g \slashletter{A} + m) /
\text{det}(\slashletter{P} + m)$, defined by the renormalized
determinant on $\mathbb{R}^d$, namely

\begin{equation}
\label{Eq_01}
\text{det}_{\text{ren}} = \exp( \Pi_2 + \Pi_3 + \Pi_4)
  \text{det}_{d+1} (1 - g S \slashletter{A}),
\end{equation}

\noindent
where

\begin{equation}
\label{Eq_02}
\text{lndet}_{d+1}
  = \text{Tr}
    \left[
      \ln(1 - gS \slashletter{A})
      + \sum^{d}_{n=1} \frac{(gS\slashletter{A})^n}{n}
    \right],
\end{equation}

\noindent
and $S = (\slashletter{P} + m)^{-1}$; $\Pi_{2,3,4}$ are the second-,
third-, and fourth-order contributions to the one-loop effective
action defined by some consistent regularization together with a
coupling constant subtraction in $\Pi_2$ when $d=4$; $\Pi_4$ is
omitted when $d < 4$ as well as $\Pi_3$ when $d = 2$. The operator
$S \slashletter{A}$ is a bounded operator on the Hilbert space
$L^2(\mathbb{R}^d, \sqrt{k^2 + m^2} d^d k)$ for
$A_{\mu} \in \cap_{n > d} L^n (\mathbb{R}^d)$, in which case it
belongs to the trace ideal $\mathcal{C}_{n}$ for $n > d$
$\left[ \mathcal{C}_n = \{A | \text{Tr}(A^{\dag} A)^{n/2} < \infty\}\right]$.
This means that $\text{det}_{d+1}$ in (\ref{Eq_02}) can be expanded in
terms of the discrete complex eigenvalues $1/g_{n}$ of the
non-Hermitian compact operator $S \slashletter{A}$:

\begin{equation}
\label{Eq_03}
\text{det}_{d+1} (1 - gS \slashletter{A})
  = \prod_{n}
    \left[ \left(1 - \frac{g}{g_{n}} \right)
      \exp
      \left(
        \sum^{d}_{k=1} \frac{\left(\frac{g}{g_{n}}\right)^{k}}{k}
      \right)
    \right].
\end{equation}

\noindent
In QED, C-invariance and the reality of $\text{det}_{\text{ren}}$ imply that the
eigenvalues of $S \slashletter{A}$ appear in quartets $\pm 1 / g_{n}$,
$\pm 1 / g^{*}_{n}$; in addition, $\Pi_3$ is absent in
(\ref{Eq_01}). It is always assumed that $m \neq 0$. The $m = 0$ case
raises nontrivial problems, and most of the above statements,
including (\ref{Eq_03}), are false in this case~\cite{A_Seiler80}. The
reader is referred to Refs.
\citen{A_Seiler80, B_Seiler82, B_Seiler81} for a review of these
basic results.

Definition (\ref{Eq_01}) of $\text{det}_{\text{ren}}$ gives back what
one gets using the Feynman rules, and so $\text{det}_{\text{ren}}$ is
what physicists mean by a determinant. Closely related to
(\ref{Eq_01}) is the Schwinger proper time representation of the
determinant provided it has a coupling constant renormalization
subtraction included. These are to be contrasted with the zeta
function definition whose field-theoretic meaning is not evident. The
modified Fredholm determinant (\ref{Eq_02}) in the definition of
$\text{det}_{\text{ren}}$ is necessary because the loop diagrams
representing $\text{Tr}(S \slashletter{A})^{n}$ for $n = 1, .., d$ are not
defined by themselves. We have summarized this basic material because
some lattice determinants in the literature, such as
$\text{det}(\slashletter{P} - g \slashletter{A} + m) = \prod_n (\lambda_n + m)$,
make no sense when extrapolated to zero lattice spacing.

\section{Analytic Results}
\label{Sec_03}

Because $\text{det}_{\text{ren}}$ has no zeros on the real $g$-axis
and $\text{det}_\text{ren} (g=0) = 1$, it is positive for real $g$. By
inspection it is an entire function of $g$, and because
$S \slashletter{A} \in \mathcal{C}_{n}$, $n > d$, it is of order $d$. This
means that for suitable constants $A(\epsilon)$, $K(\epsilon)$ and any
complex value of $g$,
$| \text{det}_{\text{ren}}| < A(\epsilon) \exp\left( K(\epsilon)
|g|^{d + \epsilon} \right)$ for any $\epsilon > 0$. For $d = 2,3$ both
the QCD and QED determinants satisfy $\text{det}_{\text{ren}} < 1$ for
real $g$ and all values of $m > 0$.\footnote{This bound is initially
derived on a lattice. There remains the technical problem of proving
that the zero lattice spacing limit of the lattice fermion determinant
coincides with $\text{det}_{\text{ren}}$ defined by (\ref{Eq_01}) and
(\ref{Eq_02}). This is partially dealt with in Refs
\citen{B_Seiler82, B_Seiler81}. A continuum proof of
$\text{det}_{\text{ren}} \leq 1$ with $g$ real for $d = 2, 3$ is
needed.}
This is known as the "diamagnetic"
bound~\cite{B_Seiler82,A_Brydges79,B_Seiler81,A_Weingarten80} which
can be interpreted in QED as a reflection of the paramagnetism of
charged fermions in a magnetic field. Therefore, in these cases
$\text{det}_{\text{ren}}$ definitely does not achieve its maximal
growth on the real $g$-axis. Already at this point we have exhausted
all that can generally be said so far about fermion determinants in
QCD and QED.

Keeping to the criteria we have stated at the end of
Sec. \ref{Sec_01}, we have only two cases to report. The first deals
with QCD's fermion determinant in the instanton background:

\begin{equation}
\label{Eq_04}
A_{\mu}(x) = A^{a}_{\mu} \frac{\tau^{a}}{2}
  = \frac{\tau^{a}}{g} \frac{\eta_{a \mu \nu} x^{\nu}}{x^2 + \rho^2}.
\end{equation}

\noindent
The $\eta$ symbols are defined in the Appendix of
Ref. \citen{A_Hooft76}; $\rho$ is a scale parameter. For $m \rho << 1$,

\begin{equation}
\label{Eq_05}
\text{lndet}_{\text{ren}}
  = \ln (m\rho) - \frac{2}{3} \ln(\mu\rho) + 2 \alpha\left(\half\right)
    + \left[ \ln(m\rho) + \gamma - \ln 2 \right] (m\rho)^2 + O(m\rho)^4,
\end{equation}

\noindent
where

\begin{equation}
\label{Eq_06}
\alpha\left(\half\right) = \frac{1}{6} (\gamma + \ln \pi)
  - \frac{\zeta'(2)}{\pi^2} - \frac{17}{72} = 0.145873,
\end{equation}

\noindent
$\gamma$ is Euler's constant and $\mu$ is the subtraction point of the
coupling constant $g$. The first three terms in (\ref{Eq_05}) were
obtained by 't Hooft~\cite{A_Hooft76}. Later, Carlitz and
Creamer~\cite{A_Carlitz79} obtained the $(m\rho)^2 \ln (m\rho)$
term. Then Kwon, Lee and Min~\cite{A_Kwon00} verified the calculation
in Ref. ~\citen{A_Carlitz79} and also obtained the remaining terms of
order $(m\rho)^2$. Note that the coefficient of the leading $\ln
(m\rho)$ term, $\mathbf{1}$, is the chiral anomaly in this case, which
should be of special interest in extrapolations of lattice
calculations of the instanton determinant.

Recently, Dunne, Hur, Lee and Min~\cite{A_Dunne} have proposed a
partial-wave WKB phaseshift method for calculating the instanton
determinant for all values of $m$. Although their predicted value for
$\alpha(1/2)$ is $0.137827$ to third order in the WKB approximation,
inclusion of higher order terms may improve this result. If so this
would raise the question of why the WKB analysis should work for other
than large values of $m$.

The inclusion of the subscript $ren$ on the determinant in
(\ref{Eq_05}) is provisional because it was not calculated following
(\ref{Eq_01}) and (\ref{Eq_02}). Instead, the $2 \alpha(1/2)$ term in
(\ref{Eq_05}) was obtained using Pauli-Villars regulator
fields\footnote{Different regulator schemes can affect the
$2\alpha(1/2)$ term in (\ref{Eq_05}). If dimensional renormalization
is used instead then

\begin{displaymath}
2 \alpha\left(\half\right) \to 2 \alpha\left(\half\right)
  - \frac{1}{3}( \ln 4\pi - \gamma) = -0.35952.
\end{displaymath}

\noindent
Eqs (13.7), (13.9) and (13.12) in Ref. \citen{A_Hooft76} relating to
this change are incorrect~\cite{B_Hooft94}.} with space-time dependent
masses followed by a change in the subtraction point to the constant
mass $\mu$ in (\ref{Eq_05}). It would be reassuring if the instanton
determinant were recalculated for $m \neq 0$ using (\ref{Eq_01}) and
(\ref{Eq_02}) or the Schwinger proper time representation without
space-time dependant masses and verify that the $2 \alpha(1/2)$ term
emerges when $m\rho << 1$.

It is easy to generalise the first two terms in (\ref{Eq_05}) to any
square-integrable self-dual background gauge field with range $\rho$
interacting with a massive fermion. Referring to Eqs. (3.5) and (3.10)
in Ref. \citen{A_Fry97} one sees by inspection that for $\text{QCD}_4$
and $\text{QED}_4$

\begin{equation}
\label{Eq_07}
\lim_{m^2 \to 0} m^2 \pdho{2}{m} \ln det_{\text{ren}}
  = \frac{1}{32\pi^2} \int d^4x \, \text{tr} \, F^2_ {\mu\nu}
    - \frac{1}{48\pi^2} \int d^4x \, \text{tr} F^2_{\mu\nu}.
\end{equation}

\noindent
The last term in (\ref{Eq_07}) arises when renormalization is carried
out by a momentum space subtraction at $p^2 = 0$. Subtracting instead
at $p^2 = \mu^2$ gives for $m\rho << 1$

\begin{equation}
\label{Eq_08}
\ln det_{\text{ren}}
  = \frac{1}{16\pi^2}[\ln(m\rho) - \frac{2}{3} \ln(\mu\rho)]
    \int d^4x \, \text{tr} F^2_{\mu\nu} + R(m^2, \mu^2),
\end{equation}

\noindent
where $R(0, \mu^2)$ is a $\mu$ and $\rho-$ independant constant that
depends of $F_{\mu\nu}$.

The second analytic result available is the fermion determinant
in $\text{QED}_{2}$ for the background field
$A_{\mu} = \epsilon_{\mu\nu} \partial_{\nu} \varphi$, where
$\epsilon_{12} = -\epsilon_{21} = 1$ and
$B = - \partial^2 \varphi$. The magnetic field $B$ is arbitrary except
that $B(r) \geq 0$ with continuous first and second derivatives and
$B(r) = 0$ for $r \geq a$ with $\int^a_0 dr \, r B^2(r) < \infty$. In
polar coordinates, $A_r = 0$, $A_{\theta} = \Phi(r) / 2\pi r$,
where $\Phi(r)$ is the total flux of $B$ out to $r$. Hence
$A_{\theta} = \Phi / 2\pi r$, $r > a$, where $\Phi$ is $B$'s total
flux. Then for the sequence of limits indicated we obtain the
following result~\cite{A_Fry03}:

\begin{equation}
\label{Eq_09}
\lim_{|e\Phi| >> 1} \lim_{ma << 1} lndet_{\text{ren}}
  = - \frac{|e\Phi|}{4\pi} \ln \left(\frac{|e\Phi|}{(ma)^2}\right)
    + O(|e\Phi|, (ma)^2|e\Phi|\ln(|e\Phi|)).
\end{equation}

The minus sign is in accord with the diamagnetic bound cited in
Sec. \ref{Sec_02}, and the coefficient of the $\ln(ma)$ term,
$e\Phi/2\pi$, is the two-dimensional chiral anomaly, assuming $e >
0$. The analysis in Ref. \citen{A_Fry03} is not sufficient to rule out
the $(ma)^2 |e\Phi| \ln |e\Phi|$ term in the remainder in
(\ref{Eq_09}); it may only be $(ma)^2 |e\Phi|$.

This result is therefore valid for a whole class of background
gauge fields in the nonperturbative region of small mass and strong
coupling. Note that the $\Pi_2$ term in (\ref{Eq_01}) is cancelled by
$det_3$ in (\ref{Eq_01}) and (\ref{Eq_02}). We believe this is a
general result and that $\Pi_4$ will be cancelled by $det_5$ in
$d = 4$.

Finally, although result (\ref{Eq_09}) is for an Abelian background field in
$d = 2$ it remains a challenge for lattice QCD fermion determinant
algorithms -- which we take to include the discretization of the Dirac
operator adopted -- to reproduce it.

\textbf{Added Note:} The authors cited in Ref. \citen{A_Dunne} have
substantially improved their result in hep-th/0410190 (``\emph{Precise
Quark Mass Dependance of the Instanton Determinant}'').

\end{document}